\documentclass[12pt]{article}
\usepackage{epsfig,amsfonts,amssymb}
\usepackage{hyperref}

\usepackage{cite}
\usepackage{cite}

\usepackage{comment}

\def\ZZZ{{\hbox{ Z\kern-1.6mm Z}}}
\def\RRR{{\hbox{ R\kern-2.4mm R}}}
\def\CCC{{\hbox{ C\kern-2.0mm C}}}
\def\zzz{{\hbox{z\kern-1mm z}}}

\newcommand{\qeq}{{\hbox{=\kern-2.3mm ? \kern.5mm }}}
\renewcommand{\qeq}{=}

\newcommand{\eps}{\epsilon}

\newcommand{\CC}{{\cal C}}

\newcommand{\OO}{{\cal O}}

\newcommand{\NN}{{\cal N}}

\newcommand{\be}{\begin{equation}}
\newcommand{\ee}{\end{equation}}
\newcommand{\ben}{\begin{eqnarray}\displaystyle}
\newcommand{\een}{\end{eqnarray}}

\newcommand{\refb}[1]{(\ref{#1})}
\newcommand{\p}{\partial}
\newcommand{\sectiono}[1]{\section{#1}\setcounter{equation}{0}}

\def\one{{\hbox{ 1\kern-.8mm l}}}
\def\zero{{\hbox{ 0\kern-1.5mm 0}}}

\newcommand{\bea}[1]{\begin{eqnarray}\label{#1} }
\newcommand{\eea}{\end{eqnarray}}

\newcommand{\eqref}{\refb}

\usepackage{bm}
\usepackage[table]{xcolor}

\def\figimaginary{

\def\JPicScale{0.8}
\ifx\JPicScale\undefined\def\JPicScale{1}\fi
\unitlength \JPicScale mm
\begin{picture}(110,60)(0,0)
\linethickness{1mm}
\multiput(10,20)(0.12,0.12){167}{\line(1,0){0.12}}
\linethickness{1mm}
\multiput(30,40)(0.12,-0.12){167}{\line(1,0){0.12}}
\linethickness{1mm}
\put(30,40){\line(0,1){20}}
\linethickness{1mm}
\put(90,20){\line(0,1){20}}
\linethickness{1mm}
\multiput(70,60)(0.12,-0.12){167}{\line(1,0){0.12}}
\linethickness{1mm}
\multiput(90,40)(0.12,0.12){167}{\line(1,0){0.12}}
\put(30,60.5){\makebox(0,0)[cc]{$\times$}}

\put(70,60){\makebox(0,0)[cc]{$\times$}}

\put(110,60){\makebox(0,0)[cc]{$\times$}}

\put(30,10){\makebox(0,0)[cc]{(a)}}

\put(90,10){\makebox(0,0)[cc]{(b)}}

\end{picture}
}

\begin{document}

\baselineskip 24pt

\begin{center}

{\Large \bf  Divergent $\Rightarrow$ Complex Amplitudes in
Two Dimensional String Theory}


\end{center}

\vskip .6cm
\medskip

\vspace*{4.0ex}

\baselineskip=18pt

\centerline{\large \rm Ashoke Sen}

\vspace*{4.0ex}

\centerline{\large \it Harish-Chandra Research Institute, HBNI}
\centerline{\large \it  Chhatnag Road, Jhusi,
Allahabad 211019, India}


\vspace*{1.0ex}
\centerline{\small E-mail:  sen@hri.res.in}

\vspace*{5.0ex}

\centerline{\bf Abstract} \bigskip

In a recent paper,
Balthazar, Rodriguez and Yin found remarkable agreement 
between the one instanton
contribution to the scattering amplitudes of
two dimensional string theory and those in the matrix model to the first subleading order.
The comparison was carried out numerically by analytically continuing the external
energies to imaginary values, since for real energies the string theory result diverges.
We use insights from string field theory to give finite expressions for the string theory
amplitudes for real energies. We also 
show analytically that the imaginary parts of the string
theory amplitudes computed this way reproduce the full matrix model results for general 
scattering amplitudes
involving multiple closed strings.

\vfill \eject

\baselineskip 18pt

\tableofcontents

\sectiono{Introduction and summary}

D-instantons represent saddle points of the path integral in second quantized string theory
and give non-perturbative contribution to the string amplitudes. The usual world-sheet
approach to computing these corrections suffer from infra-red divergences from the
boundary of the moduli spaces of Riemann surfaces. 
A concrete example of such ambiguities arose in a recent analysis of two dimensional string
theory\cite{1907.07688}. The authors of \cite{1907.07688} computed 
the D-instanton contribution to the closed 
string scattering amplitude, but could determine the final result only up to two 
undetermined
constants. Furthermore, the string theory result diverges for real energies of the external
states, and so in order to get finite results, \cite{1907.07688} evaluated the string
theory results for imaginary energies of the external states. These results were then
compared numerically 
with the results
in the dual matrix model\cite{dj,sw,gk,kr}, 
leading to the best
fit values $-1.399$ and .496 for the undertermined 
constants.\footnote{Ref.\cite{1912.07170}
extended this analysis to multi-instanton contribution. However since the analysis
was done only for the leading order terms in the expansion in powers of the string
coupling around the multi-instanton solution, 
the ambiguities of this type did not arise in
\cite{1912.07170}.}

String field theory\cite{9206084,9705241,1703.06410} 
is well poised to address the issues related to infra-red divergences
arising in string theory\cite{1703.06410,1902.00263,1702.06489}. 
It does so by drawing  insights from quantum field 
theory.
Indeed, in an 
earlier analysis\cite{1908.02782}, 
string field theory was used to determine the first of these constants 
unambiguously, leading to the value $-\ln 4$, which is within 1\% of the numerical
result $-1.399$. 
The second constant can also be fixed by this procedure\cite{appear}.
Both of these ambiguities arise from divergences in the open string channel, where
one or more of the internal open string states go on-shell or become tachyonic. In this
paper we address the problem of evaluation of the amplitude for real energies of external
states. As will be explained below, the associated 
divergences arise from internal closed string
states going on-shell. However string field theory
can be used to address  these  divergences as well.

We shall now summarize the main results, leaving the actual derivation of the results
to \S\ref{smulti} and \S\ref{simaginary}.
We begin with a review of the results of  \cite{1907.07688} in which the authors 
computed the D-instanton contribution to the  
$1\to 1$ scattering amplitude of closed string
tachyon in two dimensional string theory to first subleading order in the expansion in 
the string coupling constant $g_s$. Their result takes the form:
\ben \label{eonetoone}
A_{1\to 1} &=& 4\, \NN\, e^{-1/g_s}\,2\pi \delta(\omega_1+\omega_2) \, 
\sinh(\pi |\omega_1|)\, \sinh(\pi |\omega_2|) \nonumber \\ &&
\left[1  +  g_s\, f(\omega_1,\omega_2)+  g_s\, 
g(\omega_1) +  g_s\,  g(\omega_2) + C\,  g_s +\OO(g_s^2)\right]\, ,
\een
where $-\omega_1>0$ represents the energy 
of the incoming tachyon, $\omega_2>0$ denotes
the energy of the outgoing tachyon, $\NN$ is a normalization constant,  $C$ is a constant,
and
\ben \label{effull}
f(\omega_1,\omega_2) 
&=& 2^{-1/4}\pi^{1/2}
2^{(\omega_1^2+\omega_2^2)/2} \, {1\over \sinh(\pi|\omega_1|)\sinh(\pi|\omega_2|)}
\int_0^1 dy \nonumber \\ && 
y^{\, \omega_2^2/2} (1-y)^{1-\omega_1\omega_2} 
(1+y)^{1+\omega_1\omega_2}
\langle V_{|\omega_1|/2}(i) V_{|\omega_2|
/2}(i\, y)\rangle_D\, ,
\een
\be\label{egfull}
g(\omega)
= {2\, \pi^2} \,  {1\over \sinh(\pi|\omega|)} \int_0^\infty dt \int_0^{1/4} dx \,
\eta(it) \left( {2\pi \over \theta_1'(0|it)}\, \theta_1(2\, x|i\, t)\right)^{\omega^2/2}
\langle V_{|\omega|/2} (2\pi x) \rangle_A \, .
\ee
$\theta_1(z|\tau)$ is the odd Jacobi theta function and $\theta_1'(z|\tau)\equiv
\p_z \theta_1(z|\tau)$.
$\langle V_{|\omega_1|/2}(i) V_{|\omega_2|
/2}(i\, y)\rangle_D$ denotes the two point function on the upper half plane
of a pair of primaries in the
c=25 Liouville theory, carrying 
momenta $|\omega_1|/2$ and $|\omega_2|/2$, inserted
at $i$ and $iy$ respectively.  $\langle V_{|\omega|/2} (2\pi x) \rangle_A$ denotes the
one point function of the Liouville primary of momentum $|\omega|/2$ on an annulus
described by $0\le {\rm Re}(w)\le \pi$, $w\equiv w+2\pi \, i\, t$, with the vertex
operator inserted at ${\rm Re}(w)=2\pi x$. 
Explicit expressions for these correlation
functions can be found in \cite{1907.07688}.
In \refb{eonetoone}, 
inside the square bracket, the leading term 1 represents the product of two disk
one point functions of the closed string tachyons. The subleading terms proportional to
$g_s$ come from three sources. The term proportional to $f(\omega_1,\omega_2)$
represents the contribution from the disk two point function of closed string tachyon
vertex operators. 
The terms proportional to $g(\omega_1)$ and $g(\omega_2)$
represent the product of one point function on the disk and one point function on the
annulus of the closed string tachyon.
Finally the term involving the constant $C$ is the contribution from the product of the
two disk one point functions and the zero point function on surfaces of Euler number 1
-- a torus with a hole and a disk with two holes. The normalization constant $\NN$ was
fixed by comparison with the matrix model result to be\cite{1907.07688}:
\be\label{enorm}
\NN=-{1\over 8\, \pi^2}\, .
\ee
In the analysis of this paper we shall use \refb{enorm} without proving it.

The integral \refb{effull} 
defining $f(\omega_1,\omega_2)$ diverges from the 
$y\simeq 0$ and $y\simeq 1$ regions, while the integral \refb{egfull}
defining $g(\omega)$ 
diverges from the  $t\simeq \infty$, $x\simeq 0$ and 
$t\simeq 0$ regions. 
The divergence of $f$ from the 
$y\simeq 0$ region and that of $g$ from the $t\simeq \infty$ and 
$x\simeq 0$ regions are
associated with open string degenerations, and string field theory can be used to
get unambiguous finite results for the integrals from this 
region\cite{1908.02782,2002.04043,appear}, fixing the two
undetermined constants in the analysis of \cite{1907.07688}. 
Ref.\cite{1907.07688} dealt with the divergences associated with the 
$y\simeq 1$ and $t\simeq 0$ regions,
associated with the closed string degeneration, by working with imaginary energies.
In \S\ref{simaginary} we shall use insights from string field theory to
deal with these divergences, and describe the procedure for getting
finite results for these integrals for real energies. 
The main idea follows the one described in \cite{1607.06500}, -- to represent 
the amplitudes as Feynman diagrams of string field theory, and carry out
the integrals over momentum
variables along appropriate 
contours in the 
complex momentum 
plane. In fact we do not need to go through the analysis is detail, but simply lift
the results of \cite{1607.06500} to the case under study.\footnote{An alternative
approach is to deform the contour of integration over the
moduli of Riemann surfaces into complex plane\cite{1307.5124}.}

Once we have divergence free expressions for $f(\omega_1,\omega_2)$ and $g(\omega)$,
we can also use them to compute a general $n+1$-point amplitude in which
an incoming closed string tachyon of energy $-\omega_{n+1}>0$ scatters to $n$ 
outgoing closed string tachyons of energies $\omega_1,\cdots, \omega_n>0$. 
The result 
takes the form:
\ben\label{emulti1int}
A_{1\to n} &=& 2^{n+1}\, \NN\, e^{-1/g_s}\,  2\pi \delta(\omega_1+\omega_2+\cdots
\omega_n+\omega_{n+1}) 
\, \left\{\prod_{i=1}^{n+1} \sinh(\pi|\omega_i|) \right\} \nonumber \\ &&
\left[
1 +  g_s\, \sum_{i,j=1\atop i<j}^{n+1} f(\omega_i, \omega_j) +  g_s\, \sum_{i=1}^{n+1} g(\omega_i) 
+ C\,  g_s\right]\, .
\een
The matrix model result for this amplitude can be computed using the
ingredients given in \cite{1907.07688}. This has been described in \S\ref{smulti}
and  the result takes the form:
\ben \label{ematrix}
&&   2^{n+1}\, \NN\, e^{-1/g_s}\,  2\pi \delta(\omega_1+\omega_2+\cdots
\omega_n+\omega_{n+1}) 
\, \left\{\prod_{i=1}^{n+1} \sinh(\pi|\omega_i|) \right\} \nonumber \\ && \hskip 1in
\left[1 - i\, g_s\, \sum_{j=1}^n \omega_j \, 
\left(1 - \sum_{i=1}^n \pi\omega_i \coth(\pi\omega_i)\right)\right]\, .
\een
Equality of \refb{emulti1int} and \refb{ematrix} demands the following identity:
\ben\label{emulti4int}
&& \sum_{i,j=1\atop i<j}^{n} f(\omega_i,\omega_j) + \sum_{i=1}^n 
f(\omega_i, -\omega_1-\cdots -\omega_n) 
+ \sum_{i=1}^n g(\omega_i) + g(-\omega_1-\cdots -\omega_n)
+ C \nonumber \\
&& = - i \sum_{j=1}^n \omega_j \, 
\left(1 - \sum_{i=1}^n \pi\omega_i \coth(\pi\omega_i)\right)\, , \quad
\hbox{for $\omega_i>0$, $1\le i\le n$}\, .
\een

Note that $f$ and $g$ given in \refb{effull} and \refb{egfull} are formally
real, in the sense that the integrands appearing in their expressions are
real. On the other hand, the matrix model answer, encoded in the right hand side of
\refb{emulti4int},
is purely imaginary. This is a standard problem in the world-sheet approach to string 
theory\cite{sundborg,amano,sundborg1,9302003,9404128,9410152,berera2}, 
and
whenever an amplitude is expected to acquire an imaginary part, the corresponding 
integral over the moduli space of Riemann surfaces diverge. The relevant divergences 
in this case arise from the closed string channel, 
-- precisely those associated with the $y\simeq 1$ region for 
$f(\omega_1,\omega_2)$
and the $t\simeq 0$ regions for $g(\omega)$.\footnote{The 
divergences associated with 
the open string degeneration produces real result after being treated using string
field theory techniques of \cite{1908.02782,2002.04043,appear}.}
Our analysis in \S\ref{simaginary}, that shows how to extract finite results for both 
$f(\omega_1,\omega_2)$ and
$g(\omega)$, does contain imaginary parts. In fact the imaginary parts
turn out to have compact analytic expressions, given 
by:\footnote{I wish to thank Bruno Balthazar,
Victor Rodriguez and Xi Yin for raising the possibility of getting the imaginary 
parts of these amplitudes using unitarity cuts. Even though we did not use
this, with hindsight one can see that  the
imaginary parts  could have been obtained using the Cutkosky rules of string field 
theory\cite{1604.01783}. The approach discussed in \S\ref{simaginary} 
gives finite expressions for both, the 
real and imaginary parts of the amplitude, for real
energies.}
\be\label{efim}
f_{\rm imaginary}(\omega_1,\omega_2)
={1\over 2}\, {i\, \pi \, \omega_1 \, \omega_2} \, 
\left\{\coth(\pi\omega_1) + \coth(\pi\omega_2)\right\} 
\, {\rm sign}(\omega_1+\omega_2)\, ,
\ee
\be \label{egim}
g_{\rm imaginary}(\omega)={i\, \pi\over 2} 
\, |\omega| \left\{\omega\, \coth(\pi\omega) -{1\over \pi}
\right\}\, .
\ee
It is easy to check that the imaginary part of the left hand side of \refb{emulti4int},
computed with \refb{efim} and\refb{egim}, agrees with the right hand side of
\refb{emulti4int}. This gives an analytical proof of the imaginary part of \refb{emulti4int}.

\sectiono{Scattering involving multiple closed string tachyons} \label{smulti}

Ref.\cite{1907.07688} computed the following 
D-instanton induced amplitudes in two dimensional
string theory. The disk one point function of a closed string tachyon of  energy 
$\omega$ is given by
\be
2\, \sinh(\pi |\omega|)\, .
\ee
In this and in all the following expressions the energies will always be taken to be
outgoing, with an incoming particle regarded as carrying negative energy. In this
notation,
the disk two point function of a pair of closed string tachyons of energies $\omega_1$
and $\omega_2$ is given by
\be
4\, g_s\, \sinh(\pi |\omega_1|) \, \sinh(\pi |\omega_2|)\, f(\omega_1,\omega_2)\, ,
\ee
where $f(\omega_1,\omega_2)$ has been defined in \refb{effull}. An annulus one
point function of a closed string tachyon of energy $\omega$ is given by:
\be
2\, g_s\, \sinh(\pi |\omega|) \, g(\omega)\, ,
\ee
where $g(\omega)$ has been defined in \refb{egfull}. We shall also denote by
$C\, g_s$ the zero point function on surfaces of Euler number 1. 
Each amplitude carries an overall normalization of $\NN\, e^{-1/g_s}$ where
$\NN$ is the normalization constant given in
\refb{enorm}. Finally, the integration over the collective modes of the D-instanton
generates a factor of $2\pi\delta(\omega)$ where $\omega$ is the total energy
carried by all the external states of the amplitude. 
Multiplication by $\NN\, e^{-1/g_s}$ and the $2\pi\delta(\omega)$ factors
has to be done at the
end after taking the products of all the disconnected parts that an amplitude
may have. 

Using these results we can compute a general D-instanton induced scattering 
amplitude in which an incoming closed string tachyon of energy
$-\omega_{n+1}>0$ scatters to $n$ 
outgoing closed string tachyons of energies $\omega_1,\cdots \omega_n$. 
The leading order 
term comes from the product of $n+1$ disk one point functions, each giving a factor
of $2\, \sinh(\pi|\omega_i|)$, while the
subleading term is given by
the sum of three kinds of diagrams -- product of $(n-1)$ disk one point functions and
a disk two point function, the product of $n$ disk one point functions and an annulus
one point function and the product of $n+1$ disk one point functions and a zero point
function on a Riemann surface of Euler character 1. Therefore the result takes the form:
\ben\label{emulti1}
A_{1\to n} &=& 2^{n+1}\, \NN\, e^{-1/g_s}\,  2\, \pi\, \delta(\omega_1+\omega_2+\cdots
\omega_n+\omega_{n+1}) 
\, \left\{\prod_{i=1}^{n+1} \sinh(\pi|\omega_i|) \right\} \nonumber \\ &&
\left[
1 +  g_s\, \sum_{i,j=1\atop i<j}^{n+1} f(\omega_i, \omega_j) +  g_s\, \sum_{i=1}^{n+1} g(\omega_i) 
+ C\,  g_s\right]\, .
\een

The 
matrix model result for $A_{1\to n}$ was computed in \cite{1907.07688}. 
There it was shown that the leading
order result agrees with the leading order term in \refb{emulti1} (given by the 1
inside the square bracket) if we choose $\NN$ as in \refb{enorm}.
The subleading order result of the matrix model takes the form\cite{1907.07688}:
\be
A^{(1)} = - \delta(\omega_1+\cdots \, \omega_{n+1}) \, e^{-1/g_s} \, 2\, \pi \, i\, g_s \, 
\omega\, \sinh(\pi \omega) \, \sum_{S} (-1)^{|S|}
\int_0^{\omega_S} \, dx\,
e^{\pi (\omega - 2 x)}\, \left (x-{\omega\over 2}\right)\, ,
\ee
where
\be
\omega\equiv \omega_1+\cdots + \omega_n\, ,
\ee
$S$ is a subset of $\{1,\cdots, n\}$, $|S|$ is the number of elements
of $S$, and
\be
\omega_S = \sum_{i\in S} \omega_i\, .
\ee
After performing the integration over $x$ and summing over $S$
using the results,
\ben
&& 
\sum_S (-1)^{|S|} \, e^{-2\, \pi\, \omega_S} = \prod_{i=1}^n 
\left(1 - e^{-2\pi \omega_i}\right),
\nonumber \\ &&
\sum_S (-1)^{|S|} \, \omega_S\, e^{-2\, \pi\, \omega_S} =-
\prod_{i=1}^n \left(1 - e^{-2\pi \omega_i}\right) \, \sum_{j=1}^n \, \omega_j \, 
{e^{-2\,\pi\,\omega_j}\over 1 - e^{-2\pi\omega_j}}\, ,
\een
we get
\be\label{e6.6}
A^{(1)}=  \delta(\omega_1+\cdots \, \omega_{n+1}) \, e^{-1/g_s} \, 2\, \pi \, i\, g_s \, 
\omega\, \sinh(\pi \omega) \, 2^n \prod_{j=1}^n \sinh(\pi\omega_j)
\left[ {1\over 4\pi^2} -\sum_{i=1}^n {\omega_i \over 4\pi} \coth(\pi\omega_i)\right]\,.
\ee
Eq.~\refb{e6.6} 
would agree with the subleading order term in \refb{emulti1} if,
\ben\label{emulti3}
&& \sum_{i,j=1\atop i<j}^{n+1} f(\omega_i,\omega_j) + \sum_{i=1}^{n+1}
g(\omega_i)  + C = -i \sum_{j=1}^n \omega_j \, 
\left(1 - \sum_{i=1}^n \pi\omega_i \coth(\pi\omega_i)\right)\nonumber \\ &&
\hskip1in
\hbox{for \ $\omega_{n+1}=-\sum_{i=1}^n\omega_i$, \quad $\omega_i>0$ \ for \
 $1\le i\le n$}
\, ,
\een
or, equivalently,
\ben\label{emulti4}
&& \sum_{i,j=1\atop i<j}^{n} f(\omega_i,\omega_j) + \sum_{i=1}^n 
f(\omega_i, -\omega_1-\cdots -\omega_n) 
+ \sum_{i=1}^n g(\omega_i) + g(-\omega_1-\cdots -\omega_n)
+ C \nonumber \\
&& =- i \sum_{j=1}^n \omega_j \, 
\left(1 - \sum_{i=1}^n \pi\omega_i \coth(\pi\omega_i)\right)\, , \quad
\hbox{for $\omega_i>0$, \ $1\le i\le n$}\, .
\een

\begin{figure}
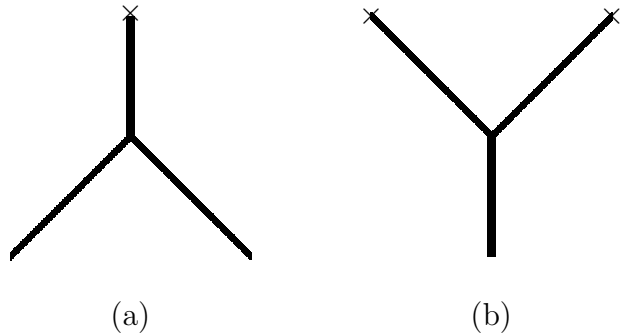

\begin{center}

\figimaginary

\caption{The Feynman diagrams contributing to the imaginary parts of $f(\omega_1,\omega_2)$ (Fig.~(a)) and $g(\omega)$ (Fig.~(b)). The thick lines
denote closed string propagators and the $\times$'s
denote the closed string one point function on the disk.
\label{figimaginary}
}
\end{center}
\end{figure}

\sectiono{Divergences to imaginary parts} \label{simaginary}

The expressions for the functions $f(\omega_1,\omega_2)$ and
$g(\omega)$, given in \refb{effull} and \refb{egfull}, 
are formally real, while the matrix model result  \refb{e6.6} is purely
imaginary. Therefore \refb{emulti4} looks wrong.
However,  the reality of $f$ and $g$ is misleading.
As in any quantum field theory, a string amplitude is expected to
acquire an imaginary part when
the energy of a subset of external states  exceeds the threshold of production
of physical intermediate states. However,  in 
the world-sheet description, 
whenever an amplitude is expected to acquire an imaginary part, the corresponding 
integral over the moduli space of Riemann surfaces 
diverges\cite{sundborg,amano,sundborg1,9302003,9404128,9410152,berera2}. 
In the present case, the imaginary parts of
$f$ and $g$, and the associated divergences in the world-sheet description,
arise from the closed string degeneration associated with the Feynman diagrams shown
in Fig.~\ref{figimaginary}. 
In order to avoid these divergences, ref.\cite{1907.07688} 
worked with imaginary energies where these divergences are absent, and
verified \refb{emulti4int} for $n=1$ in this domain numerically. 

It is however possible to work with real energies and extract finite answers 
either by deforming the contour of integration over the
moduli of Riemann surfaces into complex plane\cite{1307.5124}, or by
using string field theory
description of the amplitudes as integrals over loop momenta, and then
deforming the momentum integration contours into the 
complex plane\cite{1607.06500}. We shall illustrate this below by showing how to
get finite results for $f(\omega_1,\omega_2)$ and $g(\omega)$ 
using the string field
theory based approach, and also analytically computing the imaginary parts
of these functions for real energies.

Let us begin by analyzing the divergent part of $f(\omega_1,\omega_2)$
coming from the $y\to 1$ region of
\refb{effull}. This contribution was analyzed in eq.(2.14),
(2.15) and (2.18) of
\cite{1907.07688}. After carefully evaluating the overall normalization factors, and
accounting for the fact that we are labelling the incoming and the outgoing energies
by $-\omega_1$ and $\omega_2$ respectively, the
relevant divergent part of $f(\omega_1,\omega_2)$ takes the form:
\ben\label{edivint}
&&\hskip -.5in {1\over \sinh(\pi|\omega_1|) \sinh(\pi|\omega_2|)} \, 
\int_0^\infty dP\, \int^1 dy\, 
(1-y)^{-1 + 2 P^2 - (\omega_1+\omega_2)^2/2}
\, 2^{-2P^2 
+ (\omega_1+\omega_2)^2/2} \nonumber \\ && \hskip 2in
\CC(|\omega_1|/2, |\omega_2|/2,P)\,
\, \sinh(2\pi P) \, ,
\een
where $\CC(P_1,P_2,P_3)$ is the three point functions of primaries, carrying
momenta $P_1$, $P_2$ and $P_3$,
in $c=25$
Liouville theory\cite{9403141,9506136}, normalized as in \cite{1705.07151}.
The $y$ integral diverges\footnote{In \cite{1907.07688} this part of the
analysis was carried out for $\omega_1+\omega_2=0$, and therefore they did
not encounter this divergence.} 
for $P^2< (\omega_1+\omega_2)^2 /4$. However string field theory
tells us to replace the $y$-integral as follows\cite{1902.00263}:
\be \label{ereplace1}
\int^1 dy\, (1-y)^{-1 + 2 P^2 - (\omega_1+\omega_2)^2/2}
\Rightarrow \int^{1- a} dy\, (1-y)^{-1 + 2 P^2 - (\omega_1+\omega_2)^2/2}
+ {a^{2 P^2 - (\omega_1+\omega_2)^2/2}\over 2 P^2 - 
(\omega_1+\omega_2)^2/2-i\eps}\, ,
\ee
for any small but finite positive 
number $a$. The $i\eps$ has been introduced according to the
$i\eps$ prescription for the propagators of string field theory, and at the end we have to
take the limit $\eps\to 0^+$. In this limit, \refb{ereplace1} is an identity
for $2 P^2 - (\omega_1+\omega_2)^2/2>0$ but follows from the Feynman
rules of string field theory
for $2 P^2 - (\omega_1+\omega_2)^2/2\le 0$. It is also easy to check that the right hand
side is independent of $a$ in the $\eps\to 0^+$ limit. 
Therefore, making the replacement \refb{ereplace1}
in \refb{edivint}, one
can extract finite results for the `divergent' integral \refb{edivint}. One can even
evaluate it numerically if desired keeping the energies $\omega_1$ and $\omega_2$
real.

Let us now discuss the computation of the imaginary part of $f(\omega_1,\omega_2)$.
Since the integrand in the expression for $f(\omega_1,\omega_2)$ given in 
\refb{effull} is formally real, the only source of the imaginary part is the $i\eps$
in \refb{ereplace1}. Using the result,
\be
{1\over 2\, P^2 - (\omega_1+\omega_2)^2/2-i\eps} = {i\, \pi\over 2\, |\omega_1+\omega_2|}
\delta\left(P - {|\omega_1+\omega_2|\over 2}\right) +\hbox{real} \quad 
\hbox{for $P>0$}
\ee
in \refb{ereplace1}, and substituting this into \refb{edivint}, 
we get the following expression for the imaginary part of $f(\omega_1,\omega_2)$:
\be\label{efax1}
f_{\rm imaginary}(\omega_1,\omega_2)
={i\, \pi \over 2\, \sinh(\pi|\omega_1|) \sinh(\pi|\omega_2|)} \, 
{1\over |\omega_1+\omega_2|} \sinh(\pi|\omega_1+\omega_2|) \, \CC\left({|\omega_1|
\over
2}, {|\omega_2|\over 2},{|\omega_1+\omega_2|\over 2}\right)\, .
\ee
Using the result (see {\it e.g.} \cite{1705.07151}),
\be\label{eCid}
\CC(a,b,a+b)=\CC(a, a+b,b)=\CC(a+b,a,b)=8\, a\, b\, (a+b)\, ,
\ee
we can express \refb{efax1} as
\be \label{efax2}
f_{\rm imaginary}(\omega_1,\omega_2)
={1\over 2}\, {i\, \pi \, \omega_1 \, \omega_2} \, 
\left\{\coth(\pi\omega_1) + \coth(\pi\omega_2)\right\} 
\, {\rm sign}(\omega_1+\omega_2)\, .
\ee

Next we turn to the computation 
of $g(\omega)$. Our focus will be on extracting finite result from the 
$t\simeq 0$ region of \refb{egfull} where the integral is apparently divergent.
Comparing 
\refb{egfull} of this paper with 
eqs.(2.27)  
of \cite{1907.07688} and using the results of appendix A of \cite{1907.07688}, 
the divergent part of $g(\omega)$ from this region can be written in the form:
\ben\label{egim1}
&& {1\over \sinh(\pi|\omega|)}\, 2^{7/2}\, \pi\,  \int_0^\infty dP_1 \int_0^\infty dP_2 \ 
\CC(|\omega|/2, P_1, P_2) \, \sinh(2\pi P_1)\, \sinh(2\pi P_2)\, \nonumber \\ && \hskip .2in
\int_0^\infty ds \, \int_0^{1\over 4} dx\, 
 s^{1/2} \, \exp\left[-2\pi s \left\{ (1-2x)P_1^2 + 2x P_2^2 - x
\left({1\over 2}-x\right)\omega^2\right\}
\right]\, ,
\een
where the integration variables $s$ is related to $t$ in \refb{egfull} by $s=1/t$.
Therefore the divergence near $t=0$ now arises from the region of large $s$.
We have chosen the lower limit of $s$ integral to be 0 for definiteness, but this
has no particular significance -- we can change this to any other value at the
cost of adding a finite contribution.
Let us now change the integration variables from $(s,x)$ to
\be 
t_1 = 2\, \pi\, s\, (1-2\, x), \quad t_2 = 4\pi \, s\,  x\, .
\ee
We note further that
for $x\le 1/4$, we have $t_2\le t_1$. We shall relax this by allowing $x$
integration to run over the range $0\le x\le 1/2$ at the cost of multiplying the integrand
by a factor of $1/2$. This reduces \refb{egim1} to
\ben \label{egim2}
&& {1\over \sinh(\pi|\omega|)}\, \pi^{-1/2}\,  \int_0^\infty dP_1 \int_0^\infty dP_2 \
\CC(|\omega|/2, P_1, P_2) \, \sinh(2\pi P_1)\, \sinh(2\pi P_2)\, \nonumber \\ && \hskip .2in
\int_0^\infty dt_1 \, \int_0^\infty dt_2\, 
 (t_1+t_2)^{-1/2} \, \exp\left[-t_1 \, P_1^2 -t_2\,  P_2^2 +{t_1 t_2\over t_1+t_2} \,
 {\omega^2\over 4}
\right]\, .
\een
This integral diverges from the large $t_1,t_2$ region for $\omega> 2(P_1+P_2)$,
but there is no divergence from the region of large $t_1$ at fixed $t_2$ or vice versa.
However, as has been discussed in detail in \cite{1607.06500}, this divergence
can be attributed 
to the wrong use of Schwinger parametrization in a string field theory Feynman diagram. 
Ref.\cite{1607.06500} also
discusses how to extract finite answer from this apparently divergent integral.
Comparing eq.(2.11) of \cite{1607.06500} for $D=1$ to
the sum of eqs.(2.4) and (2.5) of \cite{1607.06500}, 
we get the replacement rule:
\ben \label{ereplace2}
&&\hskip -.3in  (4\pi)^{-1/2} \, \int_A^\infty  dt_1 \int_A^\infty dt_2 \, 
(t_1+t_2)^{-1/2} \, 
\exp\left[ {t_1 t_2\over t_1+t_2}\, M^2 - (t_1 m_1^2+t_2 m_2^2)\right]
\nonumber \\
&&\hskip-.4in \Rightarrow
\exp\left[A \left(M - m_2\right)^2 -A\, m_1^2
\right] 
\left(2\, m_2\right)^{-1} 
\left\{ M + m_1 - m_2\right\}^{-1}  
\left\{ m_1+ m_2 - M-i\eps\right\}^{-1} \, \Theta(M-m_2)\nonumber \\ &+& 
\int_{-\infty}^\infty {du\over 2\pi} \,
\exp\left[- A\left\{u^2 +m_1^2\right\} - A\left\{(u+iM)^2 
+m_2^2\right\}\right] \, 
\left(u^2 + m_1^2\right)^{-1}
\left\{(u+iM)^2 +m_2^2\right\}^{-1}\, ,\nonumber \\
\een
where $\Theta$ is the Heaviside function.
The right hand side is a finite integral. Therefore, substituting this into \refb{egim2}
after replacing $m_1,m_2,M$ by $P_1$, $P_2$, $\omega/2$ we can get finite result for
\refb{egim2}. 

The physical interpretation of \refb{ereplace2} is as follows. The left hand side
represents the contribution where we exress the two internal propagators in
Fig.~\ref{figimaginary} by their Schwinger parameter representation and carry out
the integration over the energies carried by these propagators. The right hand side
represents the result of carrying out internal energy integrals directly, by deforming
the energy integration contour to lie along the imaginary axis, and picking up
residues from the poles that the contour crosses during this deformation. 
The external energy $\propto M$ is kept real.
The first 
term on the right hand side is the residue at the pole that the contour crosses, while
the second term represents the integration along the imaginary energy axis. The
left hand side arises in the world-sheet theory, but string field theory tells us that
the right hand side is the correct one when the two differ. For $M<m_1+m_2$, both
sides are finite and
\refb{ereplace2} holds identically.

Let us now compute the imaginary part of $g(\omega)$. Again, since the integrand
for $g(\omega)$ in \refb{egfull} is formally real, the only sources of imaginary parts
of $g(\omega)$ are the $i$'s on the right hand side of \refb{ereplace2}.
The imaginary part of \refb{ereplace2} is given by:
\be \label{eim22}
i\, \pi\, \left(4\,m_1\, m_2\right)^{-1} 
\delta\left( m_1+m_2 - M\right) \, ,
\ee
where we have used the fact that the term in the last line of \refb{ereplace2}
is real -- this can be checked by making a $u\to-u$ transformation.
Using \refb{eim22} in \refb{egim2}, we get the imaginary part of $g(\omega)$:
\ben \label{egim22pre}
g_{\rm imaginary}(\omega)&=& {i\, \pi\over 
2\, \sinh(\pi|\omega|)}\,  \int_0^\infty dP_1 \int_0^\infty dP_2 \
\CC(|\omega|/2, P_1, P_2) \, P_1^{-1} \,P_2^{-1}\, \nonumber \\ && \hskip 1in
\sinh(2\pi P_1)\, \sinh(2\pi P_2) \,
 \delta\left(P_1+P_2-{|\omega|\over 2}\right)\, . 
 \een
Using \refb{eCid} we can express this as
\be\label{egsim22}
g_{\rm imaginary}(\omega)={i\, \pi\over 2} 
\, |\omega| \left\{\omega\, \coth(\pi\omega) -{1\over \pi}
\right\}\, .
\ee

Using \refb{efax2} and \refb{egsim22}, 
it is now straightforward to verify  that the imaginary
part of the left hand side of \refb{emulti4} agrees with the matrix model answer given
on the right hand side. Therefore \refb{emulti4} can now be written as an equation
involving the real parts of $f$ and $g$:
\be\label{emulti10}
\sum_{i,j=1\atop i<j}^{n} f_{\rm real}(\omega_i,\omega_j) + \sum_{i=1}^n 
f_{\rm real}(\omega_i, -\omega_1-\cdots -\omega_n) 
+ \sum_{i=1}^n g_{\rm real}(\omega_i) + g_{\rm real}(-\omega_1-\cdots -\omega_n)
+ C =0\, , 
\ee
for $\omega_i>0$. 
A class of solutions to \refb{emulti10} is provided by
\be
f_{\rm real}(\omega_i,\omega_j) = \omega_ih(\omega_j) + \omega_j h(\omega_i),
\qquad g_{\rm real}(\omega) = \omega\, h(\omega), \qquad C=0\, ,
\ee
for any function $h(\omega)$.
It will be interesting to explore whether the results
of string theory computation yields this form.

\bigskip

\noindent {\bf Acknowledgement:} I wish to thank Bruno Balthazar,  
Victor Rodriguez and Xi Yin  for useful discussions.
This work was
supported in part by the 
J. C. Bose fellowship of 
the Department of Science and Technology, India and the Infosys chair professorship.

\end{document}